\documentclass{icrc2009}
\usepackage{amssymb}
\usepackage{verbatim}
\usepackage{graphicx}   
\usepackage{caption}    
\usepackage[font=footnotesize]{subfig} 
\usepackage{fixltx2e}
\usepackage{url}

\newcommand{\shorttitle}[1]%
{\markboth{Proceedings of the 31\MakeLowercase{$^{st}$} ICRC, {\L}\'{o}d\'{z} 2009}{#1} }

\begin{document}
\title{Measurement of the atmospheric muon flux \\ with the ANTARES detector}

\author{\IEEEauthorblockN{Marco Bazzotti \IEEEauthorrefmark{1} on the behalf of the ANTARES coll.
			                     \\
\IEEEauthorblockA{\IEEEauthorrefmark{1} University of Bologna and INFN Sezione di Bologna.}}}

\shorttitle{M. Bazzotti - Measurement of the atmospheric muon flux with the ANTARES detector}
\maketitle

\begin{abstract}
ANTARES is a submarine neutrino telescope deployed in the Mediterranean Sea, at a depth of about 2500 m. It consists of a three-dimensional array of photomultiplier tubes that can detect the Cherenkov light induced by charged particles produced in the interactions of neutrinos with the surrounding medium. Down-going muons produced in atmospheric showers are a physical background to the neutrino detection, and are being studied.
In this paper the measurement of the Depth Intensity Relation (DIR) of atmospheric muon flux is presented. The data collected in June and July 2007,  when the ANTARES detector was in its 5-line configuration, are used in the analysis. The corresponding livetime is $724\,h$. A deconvolution method based on a Bayesian approach was developed, which takes into account detector and reconstruction inefficiencies. Comparison with other experimental results and Monte Carlo expectations are presented and discussed.
\end{abstract}

 
\section{Introduction}
The largest event source in neutrino telescopes is \textit{atmospheric muons}, particles
created mainly by the decay of $\pi$ and $K$ mesons originating in the interaction of cosmic rays with atmospheric nuclei. 
Although ANTARES \cite{antares_web, antares_daq, antares_OM, antares_OM1, pcoyle} "looks downwards" in order to be less sensitive to signals due to downward going atmospheric muons, these represent the most abundant signal due to their high flux.
They can be a background source because they can be occasionally wrongly reconstructed as upward going particles mimicking muons from neutrino interactions.
On the other hand they can be used to understand the detector response and possible systematic effects. In this scenario the knowledge of the underwater $\mu$ intensity is very important for any Cherenkov neutrino telescope and the future projects \cite{km3net, nemo}. Moreover, it would also provide information on the primary cosmic ray flux and on the interaction models.  \\
 
\section{Data and simulation samples}\label{samples}
The considered data sample is a selection of June and July 2007 data: only runs with good background conditions\footnote{Good run: averaged baseline rate below $120\,kHz$, burst fraction (due to biological activity) below $20\%$, muon trigger rate more than $0.01\,Hz$ and less than $10 \,Hz$.} are considered in the analysis. The livetime of the real data sample corresponds to $724\,h$.

Atmospheric muons were simulated for the 5-line ANTARES detector. The equivalent livetime corresponds to $687.5\, h$. 
The Monte Carlo programs used in the analysis are the following: 
\begin{itemize}
\item Physics generator: MUPAGE program \cite{mupage0,mupage1}. It generates the muon kinematics on the surface of an hypothetical cylinder (\textit{can}) surrounding the detector instrumented volume (see Tab. \ref{tab-mupage}). 
\item Tracking and Cherenkov light generation: KM3 program \cite{km3-prog}. 
\end{itemize}
A dedicated program inserts the background in the simulation taking it from a real run. The Monte Carlo data are then processed by the trigger software, which requires the same trigger conditions as in the real data. 

Physical information is inferred from the triggered events (both Monte Carlo and real data) by a chi-square based track reconstruction program \cite{BBfit}. Each event is reconstructed as a single muon, even if it is a muon bundle.

\begin{table}[!t]
\begin{center} 
\begin{tabular}{|c|c|c|}
\hline
\multicolumn{3}{|c|}
{\textbf{MUPAGE generation parameters}} \\
\hline 
 & Min & Max \\
\hline 
Shower Multiplicity & 1 & 100 \\
\hline 
Shower Energy (TeV) & 0.02 & 500 \\
\hline 
Zenith angle (degrees) & 95 & 180\\
\hline
\end{tabular}
\caption{Generation parameters set in the MUPAGE simulation. \label{tab-mupage}}
\end{center}
\end{table}

\section{Cut selections based on the track reconstrution algorithm}\label{section_cut} 
A chi-square reconstruction strategy \cite{BBfit} is used in the analysis.
Different fits, based on a chi-square minimization approach, are applied by the tracking algorithm:\newline
- a linear rough fit whose extracted parameters are used as starting point for the next refined fits; \newline
- a track fit which looks for a muon track; \newline
- a bright point fit which looks for a point light source as for example electromagnetic showers originated by muon interactions with matter.\newline
Particular interest in the analysis is given to the following quality parameters:
\begin{itemize}
\item \textit{nline}: number of lines containing hits used in the track fit algorithm;
\item \textit{nhit}: number of hits (single or merged) used in the track fit algorithm;
\item $\chi^2_t$: normalized chisquare of the track fit. The smaller is its value the larger is the probability that the reconstructed track belongs to a muon event;
\item $\chi^2_b$: normalized chisquare of the bright point fit. The larger is its value the larger is the probability that the reconstructed track belongs to an electromagnetic shower and not to a muon. 
\end{itemize}
\noindent

Some cuts, based on these quality parameters, are necessary to improve the purity of the data sample.

The hits used in the track reconstruction may belong only to one line (Single Line-SL event) or to more than one line (Multiple Line-ML event). 
The events detected with a single line usually have a well reconstructed zenith angle but undefined azimuth angle (if the line is perfectly vertical, the hit informations are independent from the azimuth angle of the track).
The measurement of the Depth Intensity Relation is not strictly related with the azimuth angle and for this reason single line events are also considered here.

The cuts are performed in sequence on the quality parameters of the reconstruction program. The Efficiency\footnote{The Efficiency is defined as the 
fraction of events surviving the cuts over all the reconstructed events.} and the Purity\footnote{The Purity is defined as the fraction of events with 
a zenith reconstruction error $\Delta\theta \equiv |\theta_t-\theta_m|<5^o$ over the selected events. Applicable only to MC.} of the selected data set after each cut are presented in Tab. \ref{TOT}. 
The first cut is needed in order to remove the events for which the reconstruction algorithm does not converge toward a definite value of the fitting parameters.

\begin{table}[!t]
\begin{center}
\begin{tabular}{|c|c|c|c|}
\hline
            & Efficiency(\%) & Efficiency(\%)  & Purity(\%) \\
            &  Real data     & MC data &   MC data  \\
\hline
No cut      & 100               & 100                     & 62 \\
\hline
Reconstructed track         & 99                & 99                      & 63\\
\hline 
 $nhit>5$*       & 89                & 94                      & 64 \\
\hline
$\chi^2_t<3$    & 51                & 54                      & 77 \\
\hline
$\chi^2_b>2$ & 50                & 53                      & 78 \\
\hline 
\end{tabular}
\caption{Efficiencies and Purities. The cuts are performed in sequence. 
\newline *$nhit>5$ applied only on SL events.}\label{TOT}
\end{center}
\end{table}



\section{Depth Intensity Relation}\label{DIRsec} 
In the present section the quantities are given as functions of the zenith angle $\vartheta$ obtained from the unfolded real events.

One method to derive the DIR is to compute the muon flux $I_{h_{0}}(\cos\vartheta)$ as a function of the zenith angle $\vartheta$ at a fixed vertical depth $h_0$ in the sea. Once this distribution is known, it can be transformed into the DIR using the relation \cite{lipa,gaisserbook}:
\begin{eqnarray}\label{vertical}
I_V(h) & = & I_{h_{0}}(\cos\vartheta)\cdot |\cos\vartheta| \cdot \kappa(\cos\vartheta) \\
&&[s^{-1}\cdot cm^{-2}\cdot sr^{-1}]\nonumber
\end{eqnarray}
where the subscript "V" stands for "\textit{Vertical events}" (i.e. $\cos\vartheta=-1$) and $h=h_0/\cos\vartheta$ represents the \textit{slant depth} (i.e. the distance covered in the sea water by muons, to reach the vertical depth $h_0$ with zenith angle direction $\vartheta$). In the following $h_0=2000\,m$ corresponds to the sea depth of the highest ANTARES Photomultiplier Tubes (PMTs).  
The equation \ref{vertical} is referred to as \textit{"flux verticalization"}: it transforms the muon flux $I_{h_{0}}(\cos\vartheta)$ as a function of the zenith angle $\vartheta$ at the fixed sea depth $h_0$, into the DIR $I_V(h)$, i.e. the flux  of the vertical muons as a function of the sea depth $h$.
The $|\cos\vartheta|$ and the $\kappa(\cos\vartheta)$ factors are needed in order to take into account the zenith angle dependence of the atmospheric muon flux at sea level \footnote{The sea level flux has a zenith angle dependence $\propto 1/(\cos\vartheta \cdot \kappa(\cos\vartheta))$  where the corrective factor is needed to take into consideration the Earth curvature. $\kappa(\cos\vartheta)$ can be considered equal to 1 for $\cos\vartheta < -0.5$.} \cite{lipa,gaisserbook}. 


The measured zenith distribution $N^R_m(\cos\theta_{m})$ is obtained from the track reconstruction of the selected real events.
The deconvolution procedure is a method to derive a true distribution from a measured one. In this work the goal is to transform the measured real data distribution $N^{R}_m(\cos\theta_{m})$ into its parent angular distribution $N^{R}(\cos\vartheta)$, which represents the real events crossing the can surface during the considered experimental time:   
\begin{equation}\label{deco}
N^{R}_m(\cos\theta_m)\longrightarrow (deconvolution) \longrightarrow N^{R}(\cos\vartheta)
\end{equation}
This is possible using the Monte Carlo simulations of the detector response.

Several methods to unfold data exist. The approach that has been chosen consists in an iterative method based on Bayes' theorem proposed in \cite{DAGO}.

\begin{figure}[t!]
\begin{center}
\includegraphics[width=8.4 cm] {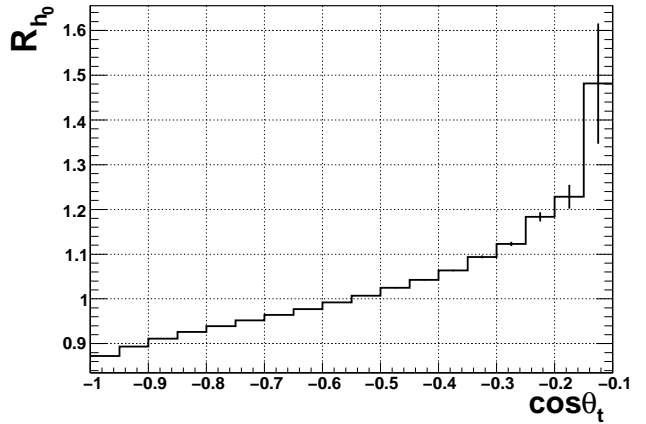}
\caption{\label{densR} $R_{h_{0}}(\cos\theta_t)$. \textbf{From Monte Carlo}. Only statistical errors are shown.}
\end{center}
\end{figure}

\begin{figure}[t!]
\begin{center}
\includegraphics[width=8.4 cm] {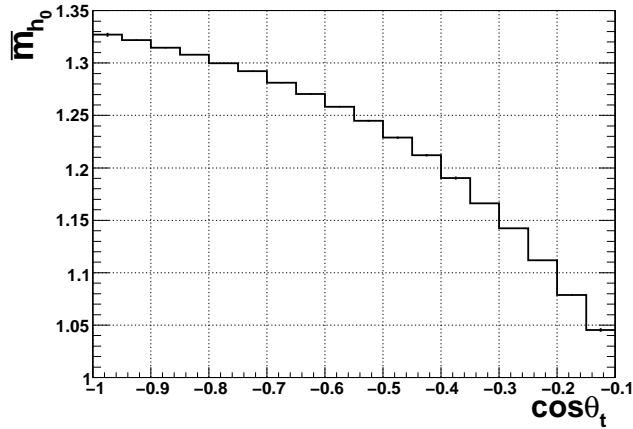}
\caption{\label{molt} Average muon event multiplicity $\overline{m}_{h_{0}}(\cos\theta_t)$ at the fixed sea depth $h_0 = 2000\,m$ for $E_\mu>20\,GeV$. \textbf{From Monte Carlo}. Only statistical errors are shown.}
\end{center}
\end{figure}

Once the distribution $N^{R}(\cos\vartheta)$ is known, it is possible to derive the atmospheric muon flux $I_{h_{0}}(\cos\vartheta)$ at the fixed depth $h_0$.
From relation \ref{vertical} the DIR can be finally written as in the following equation:
\begin{eqnarray}\label{DIRfinale}
 I_V(h) & = & \frac{N^R(\cos\vartheta)\cdot \overline{m}_{h_{0}}(\cos\vartheta)\cdot R_{h_{0}}(\cos\vartheta)}{\Delta T\cdot\Delta\Omega\cdot A_c(\cos\vartheta) } \cdot \\
 & & \cdot |\cos\vartheta|\cdot \kappa(\cos\vartheta) \;\;\;\;\; [s^{-1}\cdot cm^{-2}\cdot sr^{-1}]\nonumber
\end{eqnarray}
where the quantities in the equation are the followings:\newline
- $\Delta T=2.61\cdot10^6\,s$ is the livetime of the considered real data sample. \newline 
- $\Delta\Omega=2\pi\cdot0.05 \, sr$ is the solid angle subtended by two adjacent zenith angle bins as considered in the analysis. \newline
- $A_c(\cos\vartheta)$ is the generation can area as seen under the zenith angle $\vartheta$ (projection of a cylinder):
\begin{equation}\label{a_perp}
A_c(\cos\vartheta) = \pi R_c^2\cdot |\cos\vartheta| + 2R_c\cdot H_c\cdot |\sin\vartheta|
\end{equation}
$R_c=511\,m$ and $H_c=585\,m$ are the radius and the height of the generation can.\newline
- $N^R(\cos\vartheta)$ represents the number of muon events reaching the generation can surface during the considered experimental time $\Delta T$. \newline
- $R_{h_{0}}(\cos\vartheta)$ is a correction factor needed to get the event flux at the sea depth $h_0$ from the event flux averaged on the whole can area. $R_{h_{0}}(\cos\vartheta)$, computed from Monte Carlo ($\vartheta = \theta_t$, where $\theta_t$ is the -"true"- generated zenith angle of the Monte Carlo muon event), is shown in Figure \ref{densR}.\newline
- $m_{h_{0}}(\cos\vartheta)$ is the average muon bundle multiplicity at the fixed sea depth $h_0=2000\,m$. This quantity, computed from Monte Carlo ($\vartheta = \theta_t$), is shown in Figure \ref{molt}. \newline
- $\kappa(\cos\vartheta)\cdot \cos(\vartheta)$ are the correction factors \cite{lipa,gaisserbook} introduced in eq. \ref{vertical}.

\section{Estimation of systematic uncertainties}\label{syst}
During MC simulation several input parameters are required to define the environmental and geometrical characteristics of the detector. Some of
them are considered as sources of systematic uncertainties. 
In \cite{annarita} the effect of the variation of the following quantities on the muon reconstructed track rate is considered.
\begin{itemize}
\item Modifying by $\pm10\%$ the reference values of the sea water absorption length, an almost negligible effect on the shape of
the zenith distributions was noticed, while the absolute flux changed by $+18\%/-20\%$.
\item Decreasing and increasing the effective area of the ANTARES optical module (OM) by $10\%$ with respect to the values used in the analysis, a change of about $\pm20\%$ was observed in the muon flux.
\item The effect of the maximum angle between the OM axis and the Cherenkov photon direction allowing light collection was considered.
Moving the cut-off of this OM angular acceptance 
the rate of reconstructed tracks change of about $+35\%/-30\%$.
\end{itemize}
Summing in quadrature the different contributions, a global systematic effect of about $+45\%/-40\%$ can be
considered as an estimate of the errors produced by uncertainties on environmental and geometrical parameters.

\begin{figure}[!t]
\begin{center}
\includegraphics[width=8.4 cm] {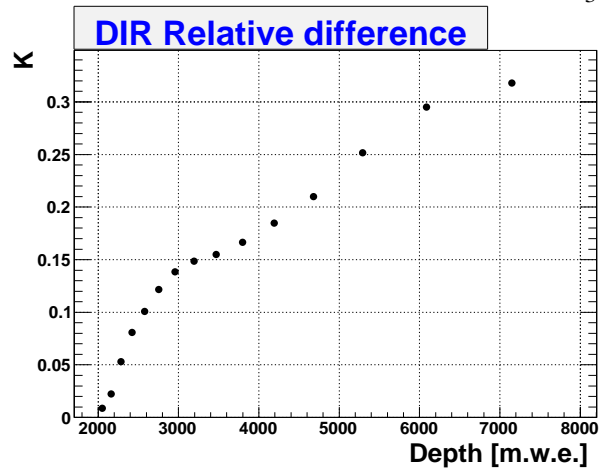}
\caption{\label{diff} Relative difference between the DIRs obtained with the defined quality cuts and without quality cuts (see eq. \ref{kfactor}).}
\end{center}
\end{figure}

The obtained results also depend on the quality cuts performed on the data set. The unfolding algorithm depends on the relative ratio of Monte Carlo and real reconstructed events used in the analysis. As seen in Tab. \ref{TOT} the selections applied to the events have slightly different effects on the two data sets.
In order to take into account this effect the unfolded DIR $I^*_V(h)$ 
has been obtained without considering any cut but the first one which eliminates not reconstructed tracks. The relative difference $K(h)$ between the two final DIRs is defined in the following equation:
\begin{equation}\label{kfactor}
K(h)=\frac{I^*_V(h) - I_V(h)}{I_V(h)}
\end{equation}
The quantity, considered as a systematic uncertainty, depends on the slant depth and is shown in Figure \ref{diff}.
This uncertainty is summed in quadrature with the systematic error estimated above. 

\begin{figure}[!t]	
\begin{center}
\includegraphics[width=8.4 cm] {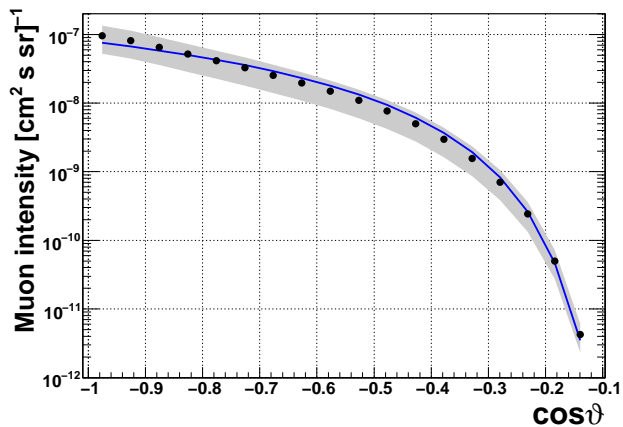}
\caption{\label{flussoNEWerr} \textbf{PRELIMINARY}. Flux of atmospheric muons for $E_\mu>20\,GeV$ at $2000\, m$ of sea depth ($I_{h_{0}}(\cos\vartheta)$) with systematic uncertainties (the statistical uncertainties are negligible). The MUPAGE simulation is superimposed.}
\end{center}
\end{figure}

In Figure \ref{flussoNEWerr} the muon flux $I_{h_{0}}(\cos\vartheta)$ ($E_\mu>20\,GeV$) at $2000$ m depth is shown. The Monte Carlo simulation from MUPAGE is also displayed. The result, as the Monte Carlo simulation, takes into account only muons with energy higher than $20\,GeV$ because muons with lower energy are not able to trigger the detector.

\begin{figure}[!t]
\begin{center}
\includegraphics[width=8. cm] {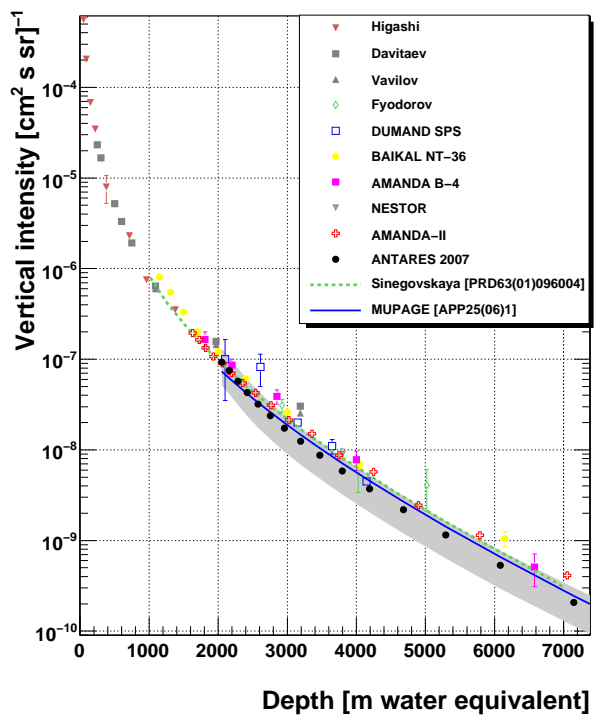}
\caption{\label{DIRallNEW} \textbf{PRELIMINARY}. Depth Intensity Relation of atmospheric muons for $E_\mu>20\,GeV$, with systematic uncertainties (the statistical uncertainties are negligible). The DIR obtained from other underwater measurements are also shown: Higashi \cite{d-higa}, Davitaev \cite{d-davitaev}, Vavilov \cite{d-vavilov}, Fyodorov \cite{d-fyodorov}, DUMAND-SPS \cite{dumand}, BAIKAL NT-36 \cite{baikal2}, NESTOR \cite{d-nestor}, AMANDA B-4 \cite{d-amandab4}, AMANDA-II \cite{d-amandaII}. The Sinegovskaya parameterization ($E_\mu>20\,GeV$) \cite{sine} and the MUPAGE simulation are superimposed.}
\end{center}
\end{figure}

In Figure \ref{DIRallNEW} the DIR $I_V(h)$ is shown together with other experimental results.
The Sinegovskaya parameterization ($E_\mu>20\,GeV$) \cite{sine} and the Monte Carlo simulation from MUPAGE are also shown. 
The ANTARES results are in reasonable agreement with the ones of the deep telescope prototypes DUMAND \cite{dumand} and NESTOR \cite{nestor} and with the data of the Baikal \cite{baikal1, baikal2} and AMANDA \cite{amanda} collaborations.

\section{Conclusions\markboth{Conclusions}{}}
The aim of the presented analysis is the measurement of the muon flux at the depth of ANTARES and the derivation of the vertical component of the atmospheric muon flux as a function of the sea depth. The goal is also to assess the performance of ANTARES in detecting muons. 
The analysis has been performed on a selection of the experimental data of June and July 2007 when the ANTARES detector was in its 5-line configuration. 

Several quality cuts have been applied on the reconstructed events in order to improve their purity, in particular concerning the zenith angle reconstruction.

An unfolding algorithm, based on an iterative method, has been applied on the selected experimental data in order to retrieve back the flux of atmospheric muons with $E_\mu>20\,GeV$ at the fixed sea depth $h_0=2000\,m$. 
The experimental DIR was finally obtained. 

The results are in good agreement, within the uncertainties, with the experimental fluxes obtained by other Cherenkov telescopes.


\end{document}